# Optimal Transcoding Preset Selection for Live Video Streaming

Zahra Nabizadeh Shahr-Babak[1], Maedeh Jamali[1], Nader Karimi[2], Shadrokh Samavi[3], Shahram Shirani[1]

[1]Electrical and Computer Engineering, McMaster University, Hamilton, Canada

[2]Electrical and Computer Engineering, Isfahan University of Technology, Isfahan, Iran

[3]Computer Science, Seattle University, Seattle, USA

**Abstract**

In today's digital landscape, video content dominates Internet traffic, underscoring the need for efficient video processing to support seamless live streaming experiences on platforms like YouTube Live, Twitch, and Facebook Live. This paper introduces a comprehensive framework designed to optimize video transcoding parameters, explicitly focusing on preset and bitrate selection to minimize distortion while respecting bitrate and transcoding time constraints. The framework comprises three main steps: feature extraction, prediction, and optimization. It leverages extracted features to predict transcoding time and rate-distortion, employing both supervised and unsupervised methods. Using integer linear programming, it identifies the optimal sequence of presets and bitrates for video segments, ensuring real-time application feasibility under set constraints. The results demonstrate the framework's effectiveness in enhancing video quality for live streaming, maintaining high standards of video delivery while managing computational resources efficiently. This optimization approach meets the evolving demands of video delivery by offering a solution for real-time transcoding optimization. Evaluation using the User Generated Content dataset showed an average PSNR improvement of 1.5 dB over the default Twitch configuration, highlighting significant PSNR gains. Additionally, subsequent experiments demonstrated a BD-rate reduction of −49.60%, reinforcing the framework's superior performance over Twitch's default configuration.

**Index Terms:** Optimization, Video Encoding, Time-Quality Tradeoff, Rate-Distortion Curve, Preset, Real-Time Application

## I. Introduction

In the current online environment, video content plays a dominant role, comprising a sizable portion of overall Internet traffic. This trend is expected to persist, leading to an even greater emphasis on video as a primary source of information consumption [1]. Live video streaming has become a ubiquitous tool for communication and entertainment in today's digital world. Platforms like YouTube Live, Twitch, and Facebook Live facilitate real-time broadcasting of events, conferences, games, and more, enabling content creators to reach a global audience instantaneously. However, ensuring a seamless and uninterrupted viewing experience for diverse viewers requires efficient video processing. In this context, fast transcoding emerges as a critical technology for successful live streaming [2].

Video transcoding involves converting video files from one format or resolution to another to ensure compatibility with various devices and network conditions. The process of transcoding is computationally intensive, requiring optimization of encoding parameters to achieve the desired balance between compression efficiency, visual quality, and processing speed [3]. At the core of this optimization process is the delicate balance between three critical metrics: bitrate, distortion, and transcoding time [4]. Various articles have proposed different methods to address these challenges or to enhance these metrics, each contributing to the field in unique ways.

Yang et al in [5] propose a novel framework for constructing efficient bitrate ladders in adaptive video streaming. This method directly predicts the optimal transcoding resolution at each preset bitrate using a temporal attentive gated recurrent network, thus eliminating the need for pre-encoding. By capturing spatial-temporal features from video clips and treating bitrate ladder estimation as a multi-task classification problem, this approach significantly reduces computational overhead while maintaining high video quality.

Wang et al. [6] propose an innovative approach to optimize video transcoding by incorporating perceptual quality of the input video. By leveraging a machine learning-based metric to detect low-quality user generated content, their quality guided transcoding framework adjusts transcoding parameters to reduce bitrate without sacrificing perceptual quality [6].

The paper [7] introduces a method to enhance immersive video streaming by optimizing the configurations of the Test Model for Immersive Video (TMIV) codec. The authors propose two Neural Network-based algorithms: a Convolutional Neural Network (CNN) and a Deep Reinforcement Learning (DRL) algorithm. The CNN algorithm addresses the configuration optimization as a regression problem, while the DRL algorithm approaches it as a decision-making problem. The algorithms are designed to maximize video quality while minimizing decoding time and bandwidth consumption.

The authors of [8] present a novel approach to enhancing the per-shot video coding framework by incorporating a complexity dimension into the optimization process. This method aims to optimize encoding parameters by considering the Rate-Distortion-Complexity (R-D-C) trade-offs. The authors propose a hyperbolic R-D-C model and apply convex hull analysis to identify the most efficient encoding parameters. By preprocessing video sequences into shots and encoding each shot with various parameters, the method filters out suboptimal versions using convex hull techniques. This approach achieves significant bitrate savings, demonstrating BD_rate [9] gains and offers flexible complexity management suitable for both professional and user-generated content.

The paper [10] introduces ViSOR, an encoding scheme that uses Video Super-Resolution (VSR) to optimize bitrate and reduce energy consumption in online streaming. By leveraging the ultrafast x265 preset, ViSOR achieves low latency but compensates for potential quality loss with client-side VSR. This method reduces bitrate and saves encoding energy while maintaining high perceptual quality.

In [11], the proposed method addresses the energy consumption of video encoding processes in HTTP adaptive streaming. By evaluating different x265 presets for 500 video sequences, the authors optimize preset selection to balance encoding time, energy consumption, and compression efficiency.

For large-scale video transcoding systems, it is common practice to apply fixed settings (such as bitrates or Constant Rate Factor (CRF)) across all videos, which fails to account for the intrinsic variability of the content. Different research efforts have shown that better performance can be achieved by optimizing transcoding parameters according to the Rate-Distortion (R-D) curves of individual videos, tailoring the process to the specific characteristics of each video [12], [13].

In this work, a sequence of video segments is considered, and the goal is to optimize the quality of the sequence while meeting the limitation of time and bitrate. To address this, in our framework, the transcoding time and video qualities are predicted for the input video. For the sequence of videos, various sets of bitrates and configuration settings are explored to achieve an optimized solution for real-time streaming. Based on the extracted features, different configurations can be selected for each video, enabling resource sharing across the sequence. To achieve this without decoding, based on [2], a model is used to predict the transcoding time, and a model is proposed to predict the R-D curve of the video in different presets. Finally, by leveraging this information and using Integer Linear Programming (ILP) [14], the optimized set of segments is selected for each sequence.

The structure of the paper is as follows: First, the optimization problem is defined in the problem statement section, where the formulation of each part is explained. In the proposed method section, the different components of our solution are described. The next section presents the experimental results, their analysis and the result of optimization processes. Finally, the conclusion of the work is provided in the last section.

## II. Problem Statement

In the domain of video transcoding, the optimization of encoding parameters plays a pivotal role in achieving desired compression efficiency, visual quality, and processing speed. At the core of this optimization lies the interplay between three key metrics: bitrate, distortion, and transcoding time. Bitrate, the amount of data required to represent a unit of video content, is influenced by factors such as the complexity of the content, the desired visual quality, and the chosen encoding parameters. Distortion, a measure of the quality loss introduced during compression, is dependent on both the bitrate and the effectiveness of the compression algorithm in preserving visual fidelity. Transcoding time, representing the duration required to encode a video sequence, is affected by factors including the computational complexity of the encoding process, the selected encoding settings, and the hardware resources available for transcoding. The plots (a) and (c) in Figure 1 show the transcoding time for two different video segments, categorized as houto and lyric types, at five different presets. As seen from the plots, the transcoding time for the same video depends on the selected preset. This dependency on the preset is also evident in the R-D behavior of the video segments. The plots (b) and (d) in Figure 1 verify this.

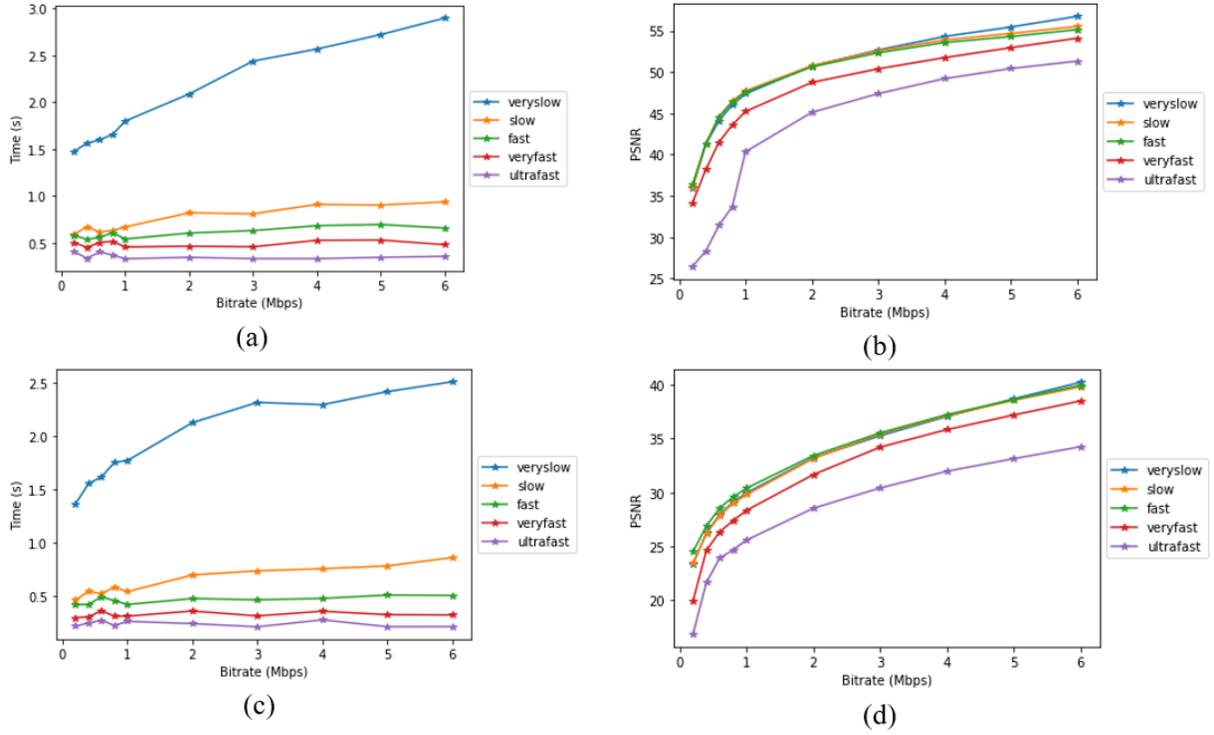

Fig. 1. The transcoding time and R-D curve of two different video types, (a, b) houto video, (c, d) Lyric video in different presets.

We formulate the transcoding optimization problem using the following equations. The first equation imposes a constraint on the total bitrate (R) across a sequence of L content segments encoded using corresponding presets (P). This constraint ensures that the total data rate of all encoded content remains within specified thresholds. The second equation introduces a constraint on the total transcoding time (T), limiting the overall processing duration for transcoding the content sequences. Restricting the processing time maintains efficient utilization of computational resources while meeting real-time encoding requirements. Lastly, the third equation formulates an objective function aimed at minimizing the total distortion (D) incurred during encoding. By seeking to minimize quality loss across all encoded content, this objective ensures that the visual fidelity of the output videos is preserved while adhering to the bitrate and transcoding time constraints. Together, these equations provide a comprehensive framework for optimizing video transcoding parameters, enabling practitioners to balance compression efficiency, visual quality, and computational resources in video transcoding applications. The following sections will describe each of these equations in detail.

## A. Total Bitrate Constraint

Equation (1) represents a constraint on the total bitrate used for encoding a sequence of $L$ segments. In video encoding, different content segments may require varying amounts of data (bitrate) to achieve desired quality levels. Each content segment is encoded using a preset (encoding configuration) denoted as $p \in \mathbb{P}$ and a bitrate denoted as $r \in \mathbb{R}$. $\mathbb{P}$ and $\mathbb{R}$ are the sets of presets and bitrates, respectively. For each content segment, there are $M = |\mathbb{P}| \times |\mathbb{R}|$ operating points where $|.|$ is the cardinality. In equation (1), the total bitrate across all $L$ content segments should not exceed a defined threshold $R_{th}$. This constraint ensures that the combined data rate of all encoded content remains within acceptable limits, which is important for the efficient use of network bandwidth or storage resources.

$$\Sigma_{i=1}^{L} \Sigma_{j=1}^{M} r_{ij} x_{ij} < R_{th} \quad x_{ij} \in \{0, 1\} \tag{1}$$

In Equation (1), $r$ represents the bitrate value, $i$ is the index for the content segment, and $j$ denotes the operating point. Each operating point, indexed by $j$, corresponds to a specific (preset, bitrate) pair used in video encoding. Therefore, $x_{ij} = 1$ indicates that the $i$-th video segment is encoded using the $j$-th operating point's specific settings, and $x_{ij} = 0$ otherwise.

## B. Total Transcoding Time Constraint

Equation (2) represents a constraint on the total transcoding time required for encoding a sequence of $L$ segments. Transcoding time ($T_{ij}$) refers to the duration it takes to encode each content segment using specific preset $p$ and a specific bitrate $r$. The total transcoding time across all $L$ content segments, represented by the summation should not exceed a specified threshold $T_{th}$. This constraint ensures that the total processing time for encoding the sequence remains within acceptable limits, which is crucial for streaming or time-sensitive applications.

$$\Sigma_{i=1}^{L} \Sigma_{j=1}^{M} T_{ij} x_{ij} < T_{th} \quad x_{ij} \in \{0, 1\} \tag{2}$$

## C. Total Distortion Minimization Objective

Equation (3) represents the objective of minimizing the total distortion incurred during encoding across the L segments of the sequence. Distortion refers to the quality loss introduced during the encoding process, typically measured using metrics such as Peak Signal-to-Noise Ratio (PSNR) or Structural Similarity Index (SSI) [15]. The objective is to find the optimal operating point that minimizes the total distortion across all L content segments, where each content segment is encoded using a specific preset p and a specific bitrate r. Minimizing distortion is essential for preserving the visual quality of the encoded content, ensuring that the encoded videos maintain high perceptual fidelity.

$$min_{(p,r)\in |\mathbb{P}|\times |\mathbb{R}|} \Sigma_{i=1}^L \Sigma_{j=1}^M D_{ij} x_{ij} \quad x_{ij} \in \{0,1\} \qquad (3)$$

In this equation, the $D_{ij}$ represents the distortion of the $i$-th content segments and the $j$-th operating point. Considering this information, finding the right balance between processing time, bitrate, and quality can be formulated as:

$$min_{(p,r)\in |\mathbb{P}|\times |\mathbb{R}|} \Sigma_{i=1}^L \Sigma_{j=1}^M D_{ij} x_{ij} \quad x_{ij} \in \{0,1\}$$

$$S.C. \quad \Sigma_{i=1}^L \Sigma_{j=1}^M r_{ij} x_{ij} < R_{th} \quad \&$$

$$\Sigma_{i=1}^L \Sigma_{j=1}^M T_{ij} x_{ij} < T_{th} \qquad (4)$$

Solving the optimization problem in Equation (4) determines the optimal bitrate and encoding preset for each video segment to achieve the best overall visual quality for the entire sequence. The challenge in solving Equation (4) is that transcoding time and distortion for a video segment depends on different parameters such as the target bitrate, the selected preset and the content of the video segment. In video streaming applications, these dependencies cannot be established without encoding the video segment at different rates and various presets as this process takes a considerable time. Therefore, building on our prior work [2] on transcoding time prediction for different presets, we propose a learning-based transcoding optimization framework. Our main contributions are:

• Proposing a learning based transcoding rate-distortion predictions using supervised and unsupervised learning approaches.

• Mapping our problem to an ILP involves using the learned transcoding rate-distortion function and transcoding time prediction to optimally assign bitrates and presets to the video segments of a video stream.

## III. PROPOSED METHOD

In live video streaming, achieving optimal trade-offs between transcoding efficiency, processing time, and output quality is crucial for delivering high-quality video content while efficiently utilizing computational resources. One of the challenges in this problem is the fact that the transcoding time and distortion of a video segment is preset dependent. Moreover, the R-D behavior of the video segments differ from each other depending on the context type. One solution to characterize these dependencies is to encode a video segment at different rates and with different presets and then solve the optimization of Equation (4). However, this 'exhaustive' approach is not optimal for live streaming applications. To address this challenge, we introduce a framework for optimizing video transcoding parameters, particularly preset selection. In this framework, we consider constraints on total bitrate and transcoding time, ensuring that resource consumption remains within acceptable bounds and live streaming is guaranteed. Additionally, we aim to minimize distortion, preserving the visual quality of the encoded content. The block diagram of our proposed framework is presented in Figure 2. This framework consists of three steps, extracting features, predicting rate distortion, and transcoding time and optimization process. In the following, these steps will be explained with more details.

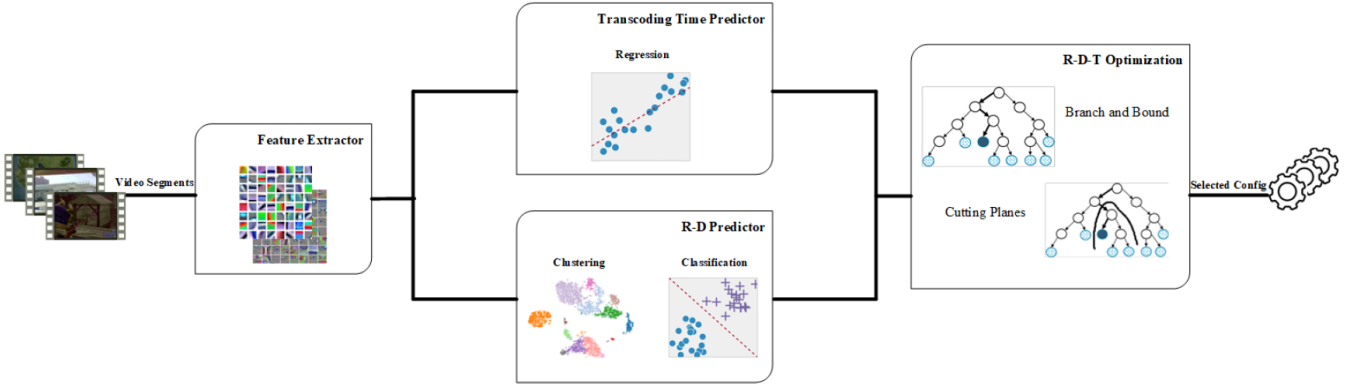

Fig. 2. The block diagram of the proposed framework.

A. Feature Extraction

Our proposed framework includes two prediction blocks, making the extracted features a crucial subblock. In this framework, features are extracted from the ingested video in the transcoder and used by two predictors to estimate transcoding time and R-D behavior. Ideal features are those that effectively predict transcoding time and R-D behavior while being quick to extract. The live streaming nature of our application limits us from using features that are complex to compute. In our study on predicting transcoding time, we extract various features categorized into frame-based, motion-based, and visual-based types, as employed in [2], for each video segment. For frame-based features, the number, and types of macroblocks, including their sizes, are crucial for describing the video's complexity. Additionally, features that describe object characteristics within the video are important for the compression process. Motion-based features, which capture the movement within the video, also play a significant role in the prediction process. Also, visual features further impact compression efficiency. We also include 'width' and 'height' in our feature vector. These features are embedded within the header and metadata of the incoming encoded stream, so extracting them incurs no extra processing time, making them ideal for our application. The complete list of features is shown in Table I, with features used for transcoding time prediction marked with a check mark the "Transcoding Time" column.

Alongside time, another key characteristic of video compression is R-D behavior. Predicting the R-D behavior of a video segment requires a dataset containing the R-D class of each segment and the features that describe it. Using these extracted features, a classifier assigns an R-D category to each video segment. Effective video classification by quality depends on integrating features that precisely capture the segment's complexity and content.

For R-D prediction, various features related to frames and motion are extracted. Some of these features overlap with those used in transcoding time prediction, while others are specifically tailored for predicting PSNR. For example, one set of features for R-D prediction includes Quantization Parameters (QP) of different frames, which influence how visual data is compressed and represented. This set includes the average QP_Y for P-frames, B-frames, and I-frames. These parameters are essential for balancing visual quality and compression efficiency, connecting to the visual representation analyzed in transcoding time feature set. Frame-based features used in this prediction are normalized to the number of frames in each video segment, providing a standardized view that facilitates comparison across videos of different lengths. Motion Vector Mean (MV_Mean) feature, which focuses on motion vectors, are key indicators of movement within the video. The inclusion of MV_Mean features highlights the importance of motion analysis in both the R-D and transcoding time prediction.

Since both sets of features for regression and classification are derived from video headers or metadata, they are well-suited for real-time applications. Table I lists all the features extracted from video segments, with features used for R-D prediction marked with "check mark" in the "R-D" column.

B. Transcoding Time Prediction using Regression Models

Transcoding time prediction involves estimating the time required to encode a video sequence based on features extracted from the ingest video. Regression models are trained using labeled data consisting of video features and corresponding transcoding times. During training, the model learns the relationship between the features and the transcoding time. Once trained, the regression model can predict transcoding time for new video sequences based on their features. The choice of regression model and the features used can significantly impact the accuracy of transcoding time prediction.

Since regression models are supervised learning models, the initial step is to generate a dataset. This involves utilizing the extracted features along with preset-based recorded transcoding times, a process which will be elaborated upon in Dataset Generation section. Subsequently, the regressors are trained using this dataset. For this purpose, the features are utilized as input, while the transcoding time serves as the output. As outlined in [2], a dedicated regressor is trained for each preset. This study employs the same regressors as [2]. By leveraging the selected regressors, transcoding times are predicted for each video sequence. Through the training of the regressors, the $T_{ij}$ value is predicted.

C. Rate-Distortion Prediction

Rate-distortion prediction entails finding the most suitable encoding parameters, such as bitrate, for each video sequence to strike a balance between compression efficiency and visual quality. This task is challenging owing to the diverse nature of videos. To tackle this problem, there are two main components, generating labels for videos and predicting distortion and bitrate. In generating labels for videos, the preparation of R-D curves and clustering are essential steps. These processes are inspired by the proposed method in [16].

In this work, different presets are considered but in [16] only the "veryfast" preset which is the default of Twitch configuration. These steps will be explained in the following.

1) Rate-Distortion Label Generation: In the realm of video transmission over the Internet, a common strategy has been to allocate uniform bitrates to all segments. However, this approach fails to consider the inherent diversity among video segments. In Figure 3, the R-D curves of four different videos show the diversity of rate distortion among different videos and different segments of one video. Some segments exhibit intricate details and extensive motion, while others are simpler in nature. Consequently, assigning identical bitrates to all segments becomes inefficient. This disparity in complexity necessitates the allocation of higher bitrates to segments with greater intricacy to maintain acceptable quality, adhering to the principles of R-D theory. At its core, R-D theory aims to represent a source using the fewest possible bits while preserving a desired level of reproduction quality [3]. This delicate balance between source fidelity and coding rate forms the essence of the rate-distortion trade-off, a crucial consideration in the design of any lossy compression system. The R-D curve offers invaluable insights into a video's behavior and facilitates the allocation of a bitrate.

For optimal transcoding, rate and distortion values are necessary. However, due to the diversity of videos, predicting precise rate and distortion values for a video segment poses a challenge. Thus, taking a learning approach can offer a solution. In this approach, the R-D curve of a large video training set is generated and clustered. To construct an R-D model for transcoding, a diverse selection of videos spanning various content genres, such as animations, lectures, news, and TV clips, was transcoded across a range of bitrates and presets. As Figure 1 (b,d) illustrates, the R-D curves for two different video types exhibit different behaviors at varying presets. Each curve in each plot corresponds to a preset, highlighting how a video segment encoded with a preset exhibits diverse PSNR values at identical bitrates. This visualization underscores the inadequacy of uniform bitrate allocation for all presets during transmission.

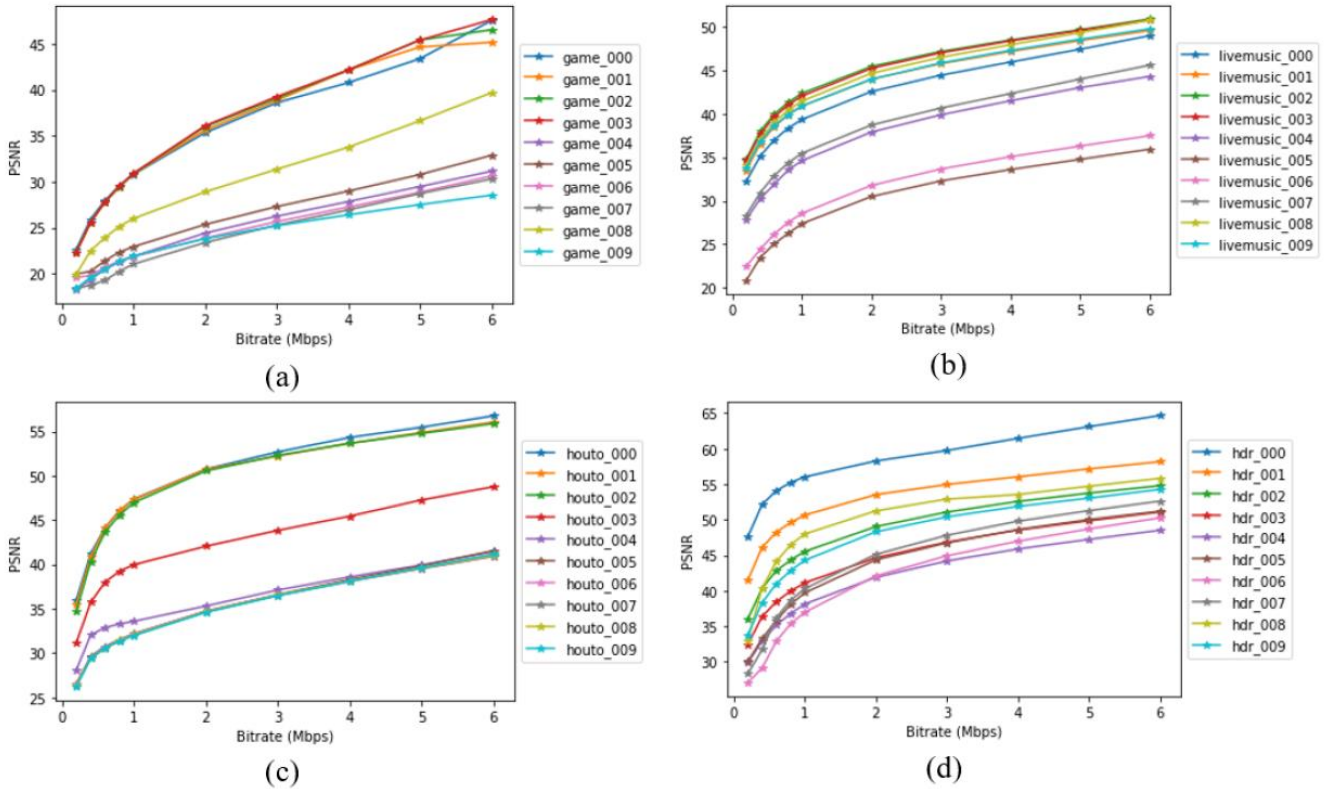

Fig. 3. R-D curve of segments of two video types, (a) game video, (b) livemusic video, (c) houto video, (d) hdr video.

Due to the diverse types and content of videos, a clustering method is employed to group similar video sequences based on their R-D curves. Clustering algorithms, such as K-means clustering or hierarchical clustering, partition the video sequences into clusters such that sequences within the same cluster are more similar than those in other clusters. Each cluster represents a set of video sequences with similar characteristics.

TABLE I Extracted Features and their usage in framework

| Feature | Feature Description | Feature Usage | |
| --- | --- | --- | --- |
| | | Transcoding Time | R-D |
| Pict_type_B | The number of B frames in the video | ✓ | ✓ |
| Pict_type_P | The number of P frames in the video | ✓ | ✓ |
| I | The number of I Macro Blocks in the video | ✓ | ✓ |
| P | The number of P Macro Blocks in the video | ✓ | ✓ |
| B | The number of B Macro Blocks in the video | ✓ | ✓ |
| S | The number of S Macro Blocks in the video | ✓ | ✗ |
| 16×16 | The number of 16x16 Macro Blocks | ✓ | ✗ |
| 16×8 | The number of 16x8 Macro Blocks | ✓ | ✗ |
| 8×16 | The number of 8x16 Macro Blocks | ✓ | ✗ |
| 8×8 | The number of 8x8 Macro Blocks | ✓ | ✗ |
| 4×4 | The number of 4x4 Macro Blocks | ✓ | ✗ |
| Sample_aspect_ratio (SAR) | Refers to the ratio of the width of a frame to its height | ✓ | ✗ |
| Ratio of skip flag for B-block | The number of skipped B-block (skipfalg=1) divided by all B-block in a GOP | ✗ | ✓ |
| Ratio of skip flag for P-block | The number of skipped B-block (skipfalg=1) divided by all P-block in a GOP | ✗ | ✓ |
| Average of QP_Y for P-frame | Summation of QP_Y of P-frames divided by the number of P-frame | ✗ | ✓ |
| Average of QP_Y for B-frame | Summation of QP_Y of B-frames divided by the number of B-frame | ✗ | ✓ |
| Average of QP_Y for I-frame | Summation of QP_Y of I-frames divided by the number of I-frame | ✗ | ✓ |
| MV | The number of Motion Vectors in the video | ✓ | ✗ |
| MV_Mean | The mean value of the magnitude of the Motion Vectors | ✓ | ✓ |
| Color_range | Determines the range of values used to represent the color information in a video frame | ✓ | ✗ |
| Color_space | Refers to the mathematical representation of the color values | ✓ | ✗ |
| Color_primaries | Refers to the standard used to define the range of colors | ✓ | ✗ |
| Color_transfer | Specifies how the color values of a video image are transformed from their original values to the values that are used for display | ✓ | ✗ |
| Width | The width of the frames of the video | ✓ | ✗ |
| Height | The height of the frames of the video | ✓ | ✗ |

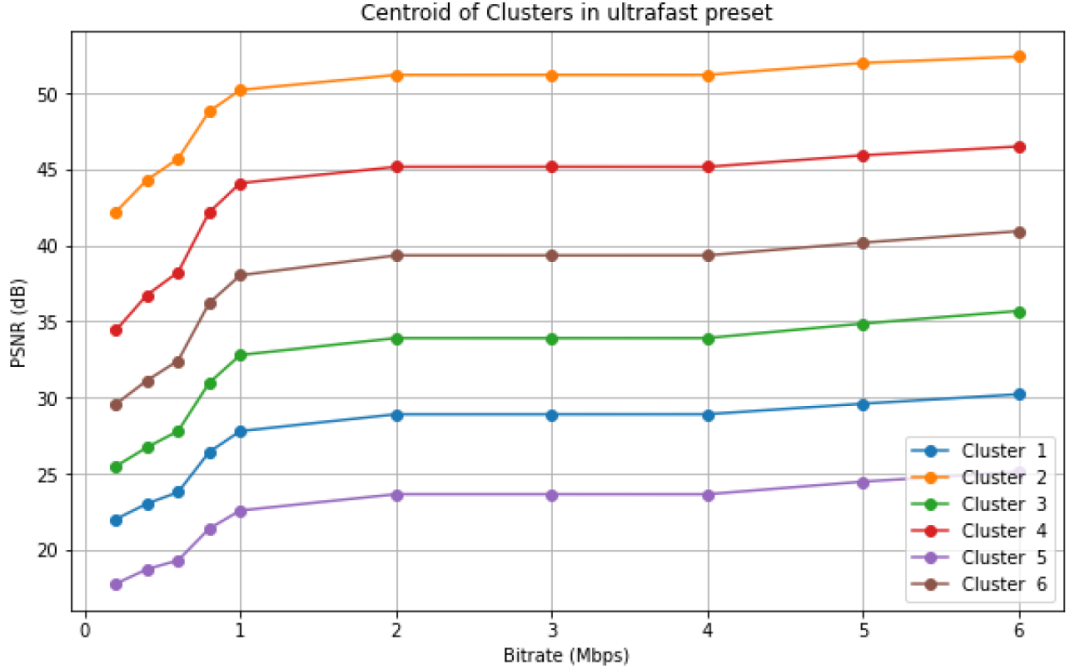

Fig. 4. The centroid of the six clusters for ultrafast preset for the video segments in dataset.

Determining the optimal number of clusters is a crucial aspect of the K-Means algorithm. For this purpose, the experimental results show that the small values result in significant gaps between the centroids, thereby restricting our ability to select an optimal bitrate tailored to specific use cases. To address this limitation, we chose to increase the number of clusters, enabling more precise bitrate selection aligned with user requirements. It is important to consider that excessively increasing the number of clusters can diminish the gaps between centroids, making it challenging to accurately distinguish between different videos. Experiments with varying numbers of clusters indicate that six clusters achieve an effective balance between precision and comprehensiveness [16].

Figure 4 highlights the centroids of these clusters for the ultrafast preset, with each centroid representing the R-D behavior for the videos within its respective cluster. The cluster centroid provides a close approximation for $D_{ij} = C_{ij}$ in Equation (4), where $C_{ij}$ represents the cluster centroid assigned to the $i$-th video segment and the $j$-th operating point (preset and bitrate).

2) Rate-Distortion Curve Classification: By utilizing the extracted features and the video cluster number, a classification model was trained. In this study, the Support Vector Machine (SVM) is chosen to determine the R-D class of each video from six categories of the R-D.

D. Rate-Distortion-Transcoding time Optimization

In the final phase of our framework, we employ ILP to solve the optimization problem, given the linearity of the objective function and the constraints defined by Equations (1) to (4). ILP is a special case of linear programming (LP) where the solution space is restricted to integer values. Our primary objective is to minimize distortion, as outlined in Equation (4), by choosing the optimal encoding preset and bitrate for each video segment. We aim to keep both the transcoding time and bitrate within their respective thresholds, as defined in Equations (1-2).

In our problem, each video segment can be transcoded at a bitrate and with a preset. We refer to this pair as the definition for that segment. In a sequence with L segments, we must select one definition for each segment to achieve minimum distortion while meeting the bitrate and transcoding time constraints. This optimization problem can be formulated as an ILP model. The relevant parameters—variables, objective, and constraints—are specified as follows:

- Decision Variables: Define $x_{ij}$ as a binary variable where $x_{ij} = 1$ indicates that definition $j$ is selected for segment $i$, and $x_{ij} = 0$ otherwise.
- Objective Function: Minimize the total distortion across all segments.
- Constraints:
    1) Each segment $i$ must have exactly one definition selected: $\sum_j x_{ij} = 1$.
    2) The total transcoding time must not exceed the threshold $T_{\text{th}}$.
    3) The total bitrate must not exceed the threshold $R_{\text{th}}$.

The ILP is formulated based on these parameters, analogous to Equations (1-4). The objective function aims to minimize the total distortion across all segments by summing the distortions $D_{ij}$ weighted by the decision variables $x_{ij}$.

The ILP is formulated based on these parameters, analogous to Equations (1-4). The objective function aims to minimize the total distortion across all segments by summing the distortions Dij weighted by the decision variables xij .

In ILP models, selecting an appropriate solver is crucial, especially for applications like real-time streaming, where the search space is expensive. For our problem, efficiency is paramount. We employ the CBC (Coin-or Branch and Cut) solver [17], [18], an open-source mixed-integer programming solver from the Computational Infrastructure for Operations Research (COIN-OR) project. CBC is specifically designed to efficiently tackle large-scale integer programming problems using advanced techniques:

• Branch and Bound: CBC employs the branch and bound algorithm to systematically explore the solution space. This method divides the problem into smaller subproblems (branching) and solves these subproblems while discarding those that do not yield better solutions (bounding).

• Cutting Planes: CBC generates and incorporates cutting planes, which are additional constraints that help to tighten the feasible region, thus improving the efficiency of the solver.

• Heuristics: CBC uses various heuristics to find good feasible solutions quickly, helping to speed up the convergence to the optimal solution.

• Presolving: Before the main solving process, CBC the performs presolving steps to simplify the problem by removing redundant constraints and fixing variable bounds, reducing the overall problem size and complexity.

various parts of the solution space simultaneously, further enhancing its performance. By leveraging these techniques, the CBC solver efficiently identifies the optimal set of encoding presets and bitrates that minimize rate-distortion while adhering to the transcoding time and bitrate constraints in real-time applications.

## IV. Experimental Results

In this section, we discuss how our proposed framework was implemented and the results. Firstly, we provide details about the platforms we used. Following that, we explain the specifics of the dataset we employed. After that the result of previous optimization works and the results obtained from our framework are reported. Additionally, we provide an analysis of the results in each part.

### A. Implementation details

The Python programming language is used for implementing the framework. FFmpeg and its extensions are used to extract features and perform transcoding in different presets. The system configuration used for our transcoding time experiments is CPU: AMD Ryzen 9 5900X @3.70 GHz. For clustering and classification method, the Intel(R) Core(TM) i7-3770 CPU @3.40GHz system is used.

### B. Dataset Generation

As regression, clustering, and classification techniques are learning-based methods, having a dataset for both training and testing is crucial. The features utilized for predicting transcoding time and R-D were discussed in Section III-A. To extract these features, the videos are segmented into two-second chunks to ensure consistent feature comparison.

In this study, we utilize the video dataset from [4], comprising eleven distinct video types such as lectures, games, sports, news, and live music, totaling 165 videos. The incoming stream bitrate to the transcoder is set to approximately 8000 Kbps and a set of target bitrates, {200,400,600,800,1000,2000,3000,4000,5000,6000} Kbps is used for transcoding with H.264 codec. Each video is segmented into two-second chunks, and the features mentioned in the Feature Extraction section are extracted. For generating transcoding time data, all videos undergo transcoding to different target bitrates with varying presets, and the transcoding time is recorded. The selected presets in this work are {veryslow, slow, fast, veryfast, ultrafast}. For the R-D aspect, the R-D curve of each video, transcoded to various target bitrates with different presets, is stored. This portion of the dataset is used for the clustering model, where the centroids of the clusters serve as target labels for the samples in the classification model.

The number of two-second chunks are 877 and the chunks in our dataset are 43850 ($877 \times 10$ (number of bitrates) $\times 5$ (number of presets). With this dataset, the three models—regression, clustering, and classification—are trained. Since the dataset does not specifically segregate training and testing data, 5-fold Cross validation is utilized for learning processes.

### C. Previous Optimization Works

In the context of optimizing the real-time video streaming process, various configuration parameters such as bitrate, preset, and rate-distortion play a crucial role. Several research articles have been published in this field, exploring the impact of these factors on streaming performance. There are no specific articles available for a direct comparison with our results. However, in this section, we report the results from several optimization studies to provide a reference range in this field.

In [5], the objective is to select the optimal resolution for each bitrate preset. This study utilizes the Inter-4K dataset [19], which includes videos at various resolutions. A deep learning model is employed to extract features, and a Bi-directional GRU is used to predict the class of bitrate and resolution. The classification achieves a maximum accuracy of 0.86, and the BD-rate is 1.21% compared to the optimal convex hull. Additionally, [5] investigates the relationship between bitrate and resolution.

In [6], the transcoding process is optimized by leveraging perceptual features to reduce bitrate without sacrificing quality. This method achieves an average bitrate reduction of up to 5%. The dataset in [6] is the same as our dataset.

In [7], the trade-off between video quality, decoding time, and bandwidth consumption is addressed in the TMIV codec. The article proposes a configuration that optimizes perceived video quality, minimizes decoding time, and reduces bandwidth consumption by transmitting an appropriate number of source views to users. In this work, an exhaustive search is used to find the optimal result, and all the results are compared with the default configuration. The utility metric, which compares the quality of videos based on their decoding time, is increased by 6%.

In [8], the per-shot encoding framework is extended by introducing complexity as an adjustable parameter, enabling encoding configurations to adapt according to resource availability and content requirements. In this approach, complexity is controlled through preset configurations, while bitrate management is achieved using Quantization Parameter or CRF. The proposed Rate-Distortion-Complexity (R-D-C) optimization model is evaluated on the BVI-1004K dataset, demonstrating up to a -19.17% BD-rate improvement compared to a conventional per-shot encoding method constrained to specific fixed presets.

This exploration demonstrates that in the optimization field of video streaming, various perspectives, parameters, and applications can be considered. Consequently, most studies in this field compare optimization results against a default or baseline. In this work, we use two baselines to evaluate and compare our results.

D. Transcoding Time Prediction

Using the generated dataset are trained for each preset using 5-fold cross validation, and the best regressor is selected among them [2]. For all presets, the Light Gradient Boosting Machine (LightGBM) regressor is selected as the best model. LightGBM is an efficient and scalable implementation of gradient boosting that uses histogram-based decision tree learning, leaf-wise growth, and techniques like Exclusive Feature Bundling (EFB) and Gradient-based One-Side Sampling (GOSS) to enhance performance. It is designed to handle large datasets and high-dimensional data with high accuracy and speed. LightGBM is widely used for various predictive modeling tasks including classification and regression [20]. The results of transcoding time prediction using LightGBM are reported in Table II. In this tabel, four evaluation metrics, Mean Squared Error (MSE), Mean Absolute Percentage Error (MAPE), Mean Absolute Error (MAE) and R-squared (R2) are reported. The first three metrics demonstrate the accuracy of the models in predicting transcoding R2 indicates how well the data points fit the model. The mean percentage error for all five presets is around 2%. For a clearer understanding, the predicted transcoding time values and the actual values for a video segment coded with different bitrates and presets are illustrated in Figure 5. In this figure, each row shows the result of one preset. To reduce the number of extracted features for real-time applications, Recursive Feature Elimination with the Cross Validation (RFECV) method the same as in [2] is implemented. During testing, the selected features are extracted solely from the videos, and the transcoding time for each preset is predicted. From the evaluation metrics presented in Table II, the MAPE of the regressor, which better illustrates the differences for each preset before and after feature selection, is reported in Table III.

TABLE II: The results of transcoding time prediction for different presets

| Model | MAE | MSE | R2 | MAPE |
|---|---|---|---|---|
| Veryslow | 0.0952 | 0.0169 | 0.9636 | 0.0318 |
| Slow | 0.0301 | 0.0017 | 0.9439 | 0.0214 |
| Fast | 0.0223 | 0.0008 | 0.8792 | 0.0185 |
| Veryfast | 0.0198 | 0.0007 | 0.8002 | 0.0180 |
| Ultrafast | 0.0149 | 0.0004 | 0.8165 | 0.0170 |
| Mean | 0.0365 | 0.0041 | 0.8807 | 0.0213 |

E. Rate-Distortion Prediction

In this section, we will explain the parameters and methods selected during the two stages of R-D prediction, R-D Label generation and classification.

1) Rate-Distortion Label generation: After extracting the R-D curves for all video segments, we apply K-means clustering with six clusters to categorize the R-D behavior of the videos. After clustering, the centroid of each cluster represents the R-D behavior for that cluster and the videos within it. The cluster number is then assigned as the R-D class for each video. Details of the clustering and curve fitting process are explained in [16]. While [16] uses only one preset, this paper applies the method to videos encoded with various presets. Figure 6 illustrates the clustering results for these presets. Each plot in this figure displays the range of quality changes and the clustering for each preset. As shown in Figure 6, videos encoded with the ultrafast preset exhibit lower PSNR values compared to those encoded with the very slow preset. The highest PSNR value for the ultrafast preset is approximately 60 dB, which is lower than that achieved by the other presets. For each cluster, a centroid is depicted, like the one shown in Figure 4. This centroid represents the R-D curves within the corresponding cluster. To identify the most appropriate curve for each cluster, we employ curve fitting techniques akin to those described in [16]. Figure 7 presents the centroids of the six clusters for different presets after curve fitting. Each plot shows the centroid of one cluster number for different presets. As illustrated in the figure, the ultrafast preset is notably distinct from the other presets, reflecting a significant difference in video quality.

TABLE III: Transcoding time of a video segment with different presets after feature selection

| Model | No Features | MAPE After | MAPE Before |
|---|---|---|---|
| Veryslow | 17/22 | 0.0320 | 0.0318 |
| Slow | 19/22 | 0.0210 | 0.0214 |
| Fast | 19/22 | 0.0184 | 0.0185 |
| Veryfast | 17/22 | 0.0177 | 0.0180 |
| Ultrafast | 19/22 | 0.0171 | 0.0170 |
| Mean | 18.2/22 | 0.0212 | 0.0213 |

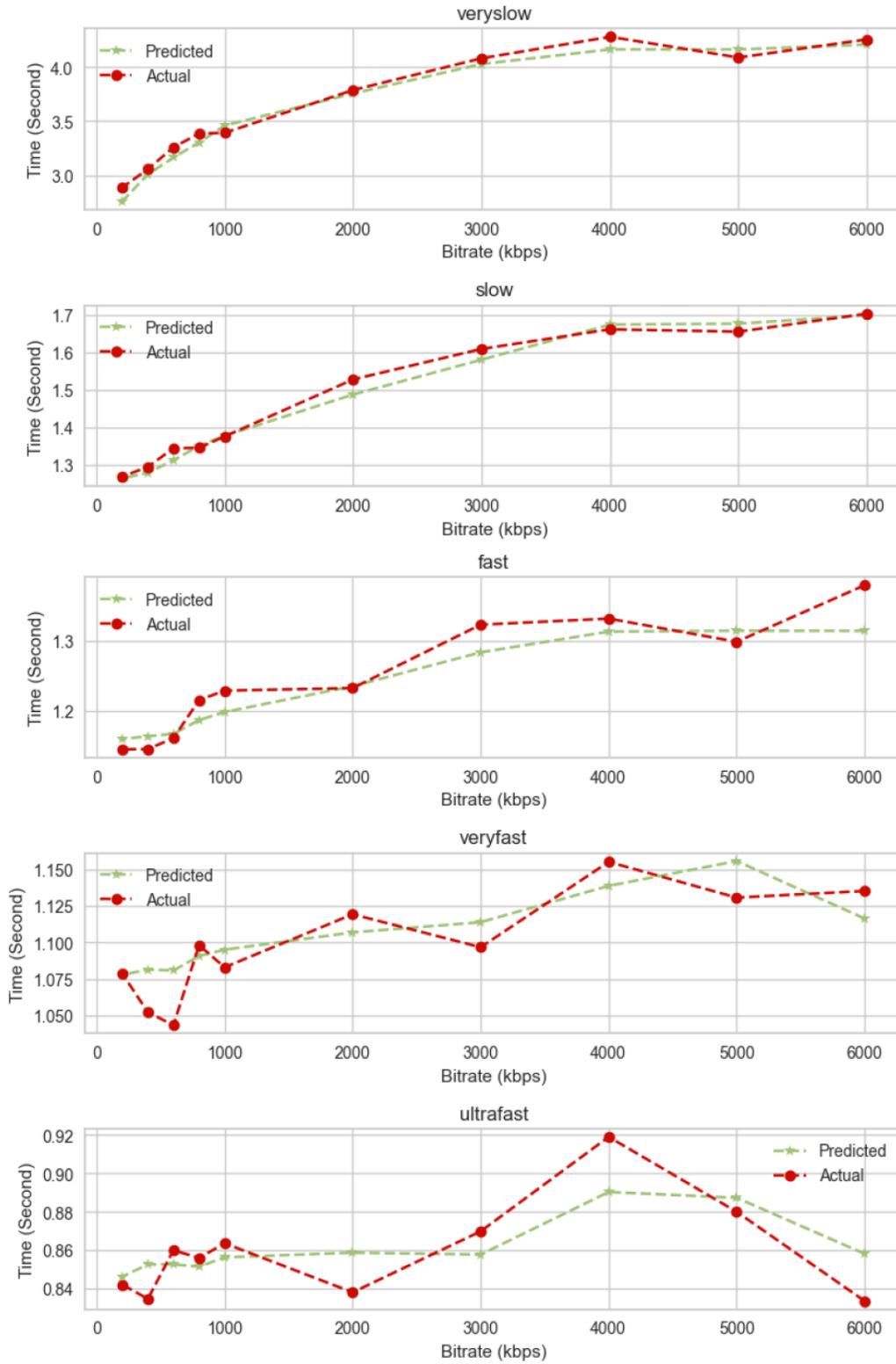

Fig. 5. The results of the predicted values and the actual values of one video for different bitrates and presets after feature selection.

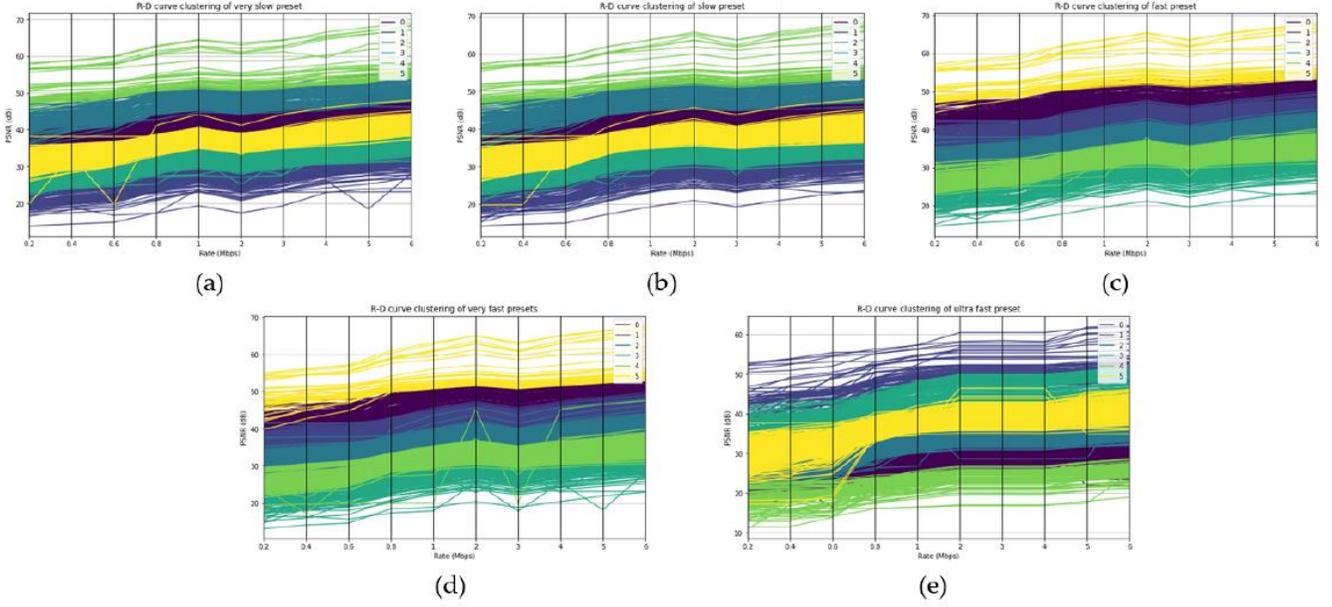

Fig. 6. R-D curve clustering of different presets using KMeans clustering, (a) veryslow preset, (b) slow preset, (c)fast preset, (d) veryfast preset, (e) ultrafast preset

2) Rate-Distortion Class Prediction: The SVM classification model is trained to predict the class (cluster) of each video at each preset. We consider several kernels such as linear, RBF, and polynomial (2nd order) and learn the SVM using these kernels. In order to have more accurate results, 5-fold cross validation is used during training. Based on achieved results in Table IV, the polynomial kernel 2nd order is selected as the final kernel in classification method. The achieved accuracy in polynomial classification is 0.73. By identifying the class of each video, its centroid R-D curve determines its rate and distortion characteristics.

F. Optimization Result

After predicting all the required information, including transcoding time and R-D curve, we proceed to select the optimized preset and bitrate that minimize distortion. In our experiments, the value for L is considered six, and the number of definitions for each segment (M) is fifty (10 bitrates × 5 presets). According to the block diagram in Figure 2, the prediction of the transcoding time and R-D could be done in parallel.

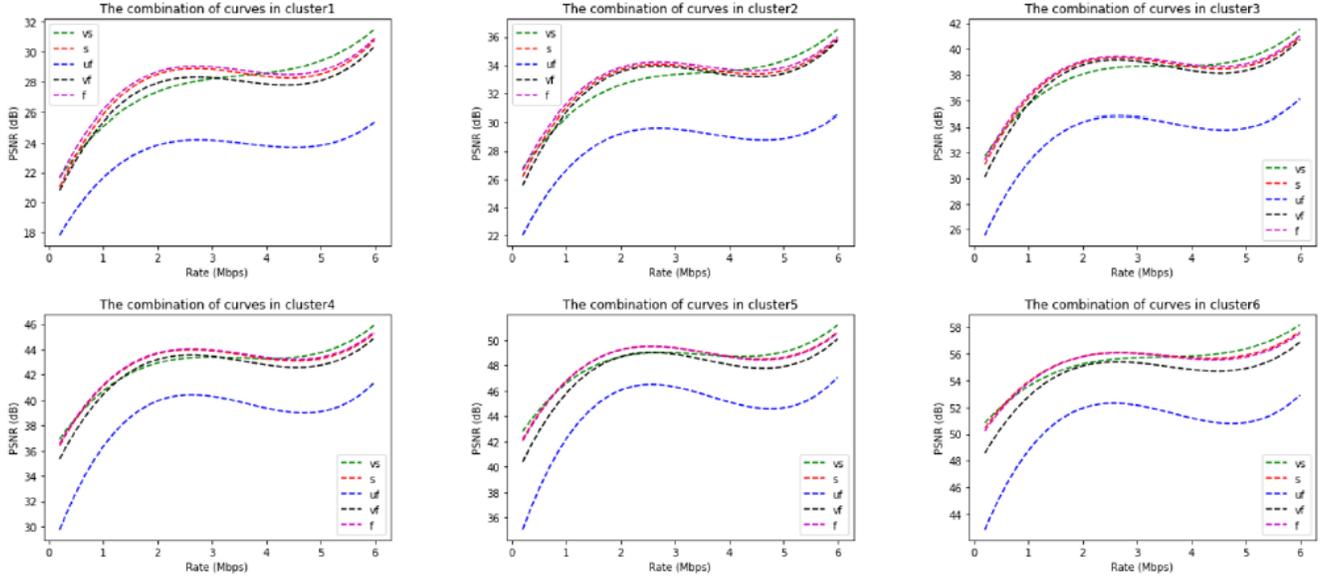

Fig. 7. Combination of cluster centroid fitted curves for different presets in each cluster.

TABLE IV: SVM classification using different kernels

| Kernel | Accuracy |
|---|---|
| Linear | 0.510 |
| RBF | 0.601 |
| Polynomial | 0.730 |

To set $T_{th}$, the maximum processing time for predicting transcoding time and R-D are considered. The processing time for transcoding time prediction is 0.02 seconds, and for R-D prediction, it is 0.0002 seconds. Therefore, the prediction time is 0.02 seconds. Since the video segments are two seconds, considering the processing time for prediction (0.02 seconds) and optimization (0.02 seconds), the thresholds for time can be defined as 11.76 ($6 \times (2 - 0.04)$) seconds. In our experiments, it is set to 11 seconds. To define the threshold for bitrate, the bandwidth of video streaming platforms is explored. Bandwidth used by platforms like Twitch1 for video streaming, is between 3000 kbps and 6000 kbps, the bitrate is 5000 kbps. Based on this, the threshold is defined as 30000 ($6 \times 5000$). By considering these values, the ILP algorithm is applied to six randomly selected chunks from a total of 877 chunks. This process is repeated 877 times, and for each run, the sum of PSNR, transcoding time, and bitrate for the selected options for each segment is recorded. For these selected chunks, the sum of PSNR, transcoding time, and bitrate is also recorded for the scenario where all segments are encoded with a 5000 kbps bitrate using the fast (baseline1) and veryfast (baseline2) presets. The fast preset is the middle preset in our selection set, while veryfast is the default preset for platforms like Twitch 2.

In Table V, the average PSNR for two scenarios—selecting options with ILP and the default options are reported. The ILP output shows a significant improvement in PSNR, increasing it by 9.45 dB compared to the veryfast preset (approximately 1.5 dB per segment). Similarly, when compared to the fast preset, the ILP output results in an increase of 5.92 dB in PSNR (approximately 1 dB per segment).

G. Bitrate and Transcoding Time Limitation

In our work, there are two constraints, bitrate and transcoding time, that we simulate to reflect different streaming scenarios. By changing each of these parameters, the selected operating point for each video segment varies. In this section, we consider the effect of these two parameters. To evaluate the compression efficiency between our proposed method and the baseline2 method, we computed the BD-rate, a widely used metric for comparing the rate-distortion performance of video codecs. The BD-rate quantifies the average percentage difference in bitrate required by the two codecs to achieve the same quality (measured by PSNR). For this purpose, by considering the same bitrate threshold ($R_{th}$) for the ILP and the baseline2, the PSNR of the encoded segments with these two configurations was recorded.

The considered values for $R_{th}$ are $6 \times \{200, 400, 600, 800, 1000, 2000, 3000, 4000, 5000, 6000\}$ Kbps. The $T_{th}$ is fixed in this experiment (11 seconds). Given these settings, each experiment was run 877 times, and the mean average PSNR was calculated for both methods. The R-D curves of the two methods are shown in Figure 8. The calculated BD-rate for this experiment is -49.60%.

TABLE V: The results of the ILP and other two baselines

| Metric | ILP | Baseline1 | ILP | Baseline2 |
| --- | --- | --- | --- | --- |
| PSNR | 253.38 | 247.46 | 253.89 | 244.44 |

To demonstrate the effect of time constraints, the bitrate constraint was fixed while the time constraint was varied. By altering the time limitation, the distribution of selected presets across diverse types of videos was recorded. In these experiments, the transcoding time limitation was set to [11, 8, 5, 3] seconds. This approach helps to identify which types of videos are more complex. For example, Figures 9 and 10 illustrate the distribution of selected presets for two types of videos across various time constraints. These figures enable inference on the influence of video content and time constraints on preset selection. It is evident how changes in transcoding time constraints affect the preset distribution for these two types of videos, highlighting the importance of selecting the optimal operating point, especially in real-time applications.

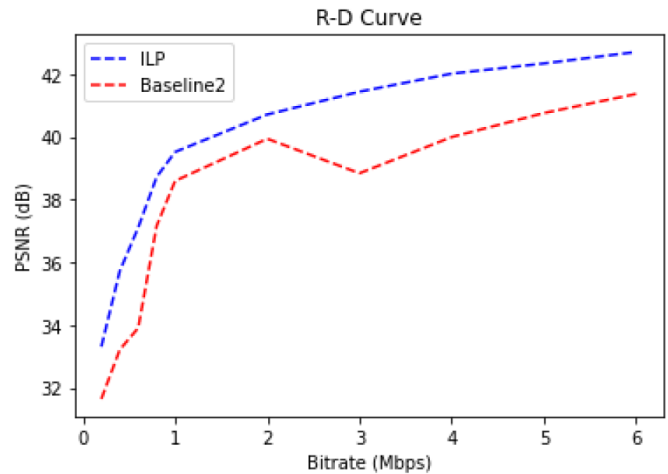

Fig. 8. The R-D curve of ILP and baseline2 method.

Furthermore, the results suggest that sports videos are more complex than music videos, evidenced by the lack of response when the time limitation is set to 3 seconds. This complexity is corroborated by the average extracted motion vectors, where the mean magnitude for sports videos is 711,971.4, compared to 547,081.9 for music videos.

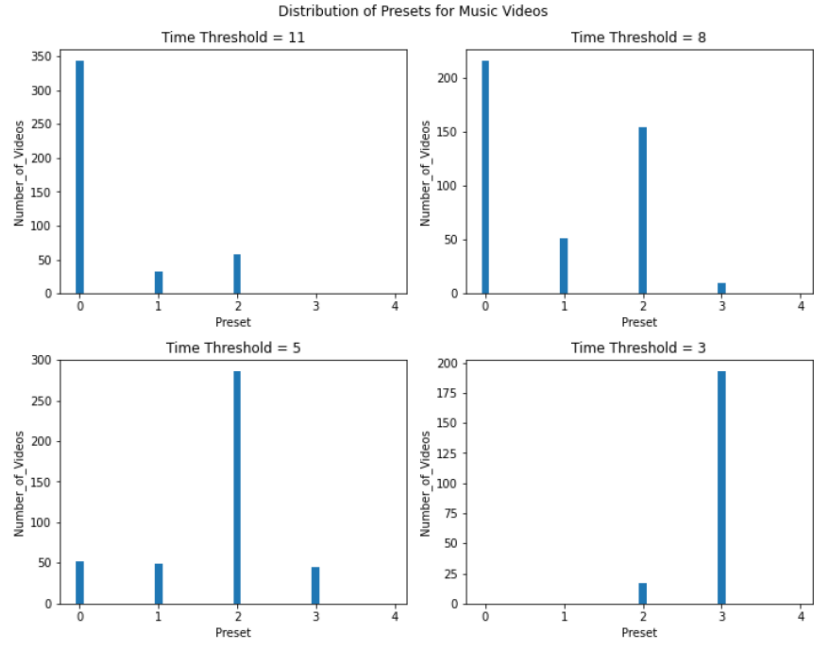

Fig. 9. Distribution of Presets for Music Videos.

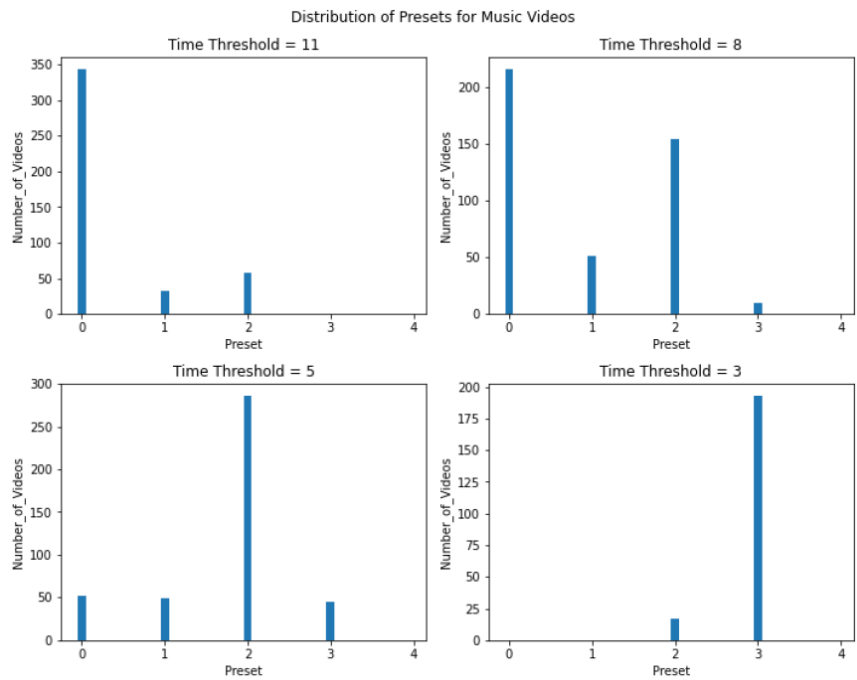

Fig. 9. Distribution of Presets for Music Videos.

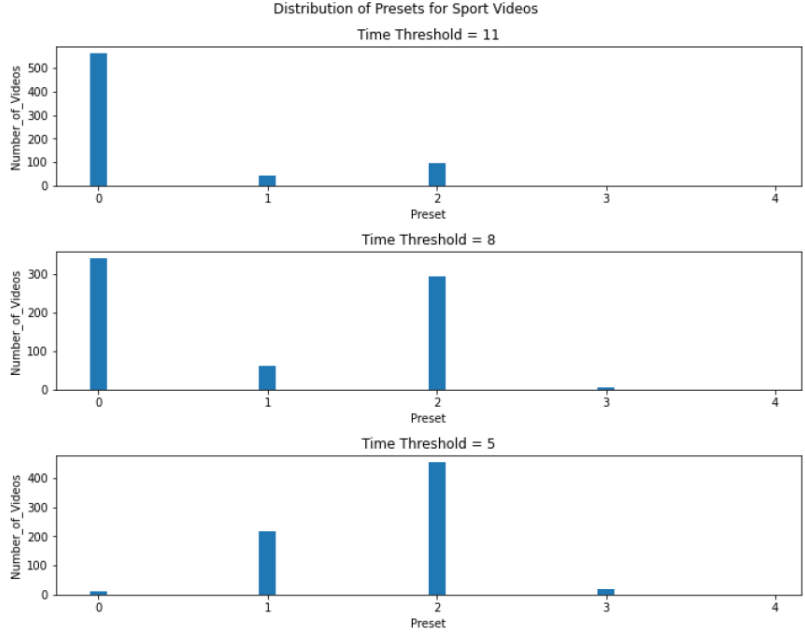

Fig. 10. Distribution of Presets for Sport Videos.

# V. Conclusion

In this work, we propose a comprehensive framework to optimize video transcoding parameters, with a focus on selecting presets and bitrates to minimize distortion while adhering to constraints on bitrate and transcoding time. By utilizing extracted features for both transcoding time and R-D prediction, our approach employs linear programming to determine the most efficient sequence of presets and bitrates for video segments in real-time applications. Although predicting R-D is a challenging problem, our approach achieves an accuracy of 73%. Our experimental results demonstrated significant improvements in PSNR compared to both baseline configurations. The ILP algorithm, when applied across multiple video segments, consistently outperformed both baselines, achieving substantial gains in video quality. Specifically, the ILP output increased the PSNR by 9.45 dB over baseline2 and by 5.92 dB over baseline1, which corresponds to approximately 1.5 dB and 1 dB improvements per segment, respectively. These results also highlight the impact of transcoding time limitations on the diversity of preset selection and underscore the importance of making informed preset choices.

These findings highlight the effectiveness of our proposed method in enhancing video quality for live streaming platforms. By optimizing encoding parameters, our framework ensures high-quality video delivery while efficiently managing computational resources. This work contributes to the ongoing efforts to improve live video streaming, providing a robust solution for real-time video transcoding optimization.